# Improved axisymmetric lattice Boltzmann scheme

Q. Li, Y. L. He, G. H. Tang, and W. Q. Tao

National Key Laboratory of Multiphase Flow in Power Engineering, School of Energy and Power Engineering, Xi'an Jiaotong University, Xi'an, Shaanxi 710049, China

This paper proposes an improved lattice Boltzmann scheme for incompressible axisymmetric flows. The scheme has the following features. First, it is still within the framework of the standard lattice Boltzmann method using the single-particle density distribution function and consistent with the philosophy of the lattice Boltzmann method. Second, the source term of the scheme is simple and contains no velocity gradient terms. Owing to this feature, the scheme is easy to implement. In addition, the singularity problem at the axis can be appropriately handled without affecting an important advantage of the lattice Boltzmann method: the easy treatment of boundary conditions. The scheme is tested by simulating Hagen-Poiseuille flow, three-dimensional Womersley flow, Wheeler benchmark problem in crystal growth, and lid-driven rotational flow in cylindrical cavities. It is found that the numerical results agree well with the analytical solutions and/or the results reported in previous studies.

Ⅰ. INTRODUCTION

Because of its kinetic nature and distinctive computational features, the lattice-Boltzmann (LB) method, which originates from the lattice-gas automata (LGA) method [1], has been developed into a very attractive alternative to conventional numerical methods. In the LB method, instead of solving the macroscopic governing equations, the discrete Boltzmann equation with certain collision models, such as the matrix model [2, 3], Bhatnagar-Gross-Krook (BGK) model [4-7], multiple-relaxation-time (MRT) model [8-13], and the two-relaxation-time (TRT) model [14-16], is solved to simulate fluid flows and model physics in fluids.

In the literature, the main advantages of the LB method are summarized as follows [17]: (i) non-linearity (collision process) is local and non-locality (streaming process) is linear, while in the Navier-Stokes equation the convective term $u\nabla u$ is non-linear and non-local at a time; (ii) streaming is exact; (iii) complex boundary conditions can be easily formulated in terms of elementary mechanics rules; (iv) fluid pressure and the strain tensor are available locally; (v) nearly ideal amenability to parallel computing (low communication/computation ratio). Owing to these advantages, in the past two decades the LB method has been successfully applied to various flow problems in science and engineering [18-24]

In recent years, the LB method for axisymmetric flows has attracted much attention. It is known that LB simulations of axisymmetric flows can be handled with a standard three-dimensional (3D) LB model. However, such a treatment does not take the advantage of the axisymmetric property of the flow: 3D axisymmetric flows are two-dimensional (2D) problems in a cylindrical coordinate system. To make use of this property, much research has been conducted. The first attempt was made by Halliday *et al*. [25]. The basic idea of Halliday *et al*.'s method is to incorporate spatial and velocity dependent source terms into the microscopic evolution equation to mimic the additional axisymmetric contributions in cylindrical coordinates. Following Halliday *et al*.'s work, Peng *et al*. [26] proposed a hybrid LB model for incompressible axisymmetric thermal flows by solving the azimuthal velocity and the temperature with a second-order center-difference scheme. Nevertheless, it was later found that Halliday *et al.*'s model fails to reproduce the correct hydrodynamic momentum equation due to some missing terms. After considering these terms, Lee *et al.* [27] developed a more accurate axisymmetric LB model. Reis and Phillips [28] have also presented a modified version of Halliday *et al.*'s model by deriving the source terms in a different manner. The modified model was subsequently validated with several numerical tests [29].

In the above models, some complex differential terms were introduced into the second-order source term due to the discrete effects on the first-order source term (identical to a forcing term). These complex terms may introduce some additional errors and do harm to the numerical stability. He *et al.* [30, 31] have pointed out the trapezium rule is necessary for the integration of a forcing term to avoid the spurious effects in the recovered macroscopic equations. By using a new distribution function to eliminate the implicitness resulting from the trapezium rule, it can be found that a factor dependent on the relaxation time will be included in the forcing term and the macroscopic variables should be redefined [32]. Following this strategy, Premnath and Abraham [33] devised a LB scheme for axisymmetric multiphase flows. The scheme was extended to axisymmetric two-phase flows with large density ratio in Ref. [34]. Similarly, Zhou [35] recently proposed a simplified axisymmetric LB model by adopting a centered scheme to simplify the source term.

Besides the above-mentioned models, an axisymmetric LB method based on the vorticity-stream-function equations of incompressible axisymmetric flows has also been developed [36, 37]. In this method, distribution functions $rf_\omega$ and $f_\psi$ for $\omega$ and $\psi$ are adopted, where $\omega$ and $\psi$ are the vorticity and the stream function, respectively, and $r$ is the coordinate in the radial direction. The radial velocity $u_r$ and axial velocity $u_z$ are obtained from $u_r = (\partial_z \psi)/r$ and $u_z = -(\partial_r \psi)/r$ with a second-order center-difference scheme, where $z$ is the coordinate in the axial direction. The authors pointed out that the main drawback of the method is the difficulty in treating boundary vorticity. In this regard, an important advantage of the LB method, the easy treatment of boundary conditions, may be lost. In a recent paper [38], an axisymmetric kinetic BGK model of single-particle density distribution function $f$ has been derived from the continuous Boltzmann equation in cylindrical coordinates. Due to a term in the kinetic model, a new distribution function $rf$ was employed to replace $f$ when devising an axisymmetric LB model. The source term of the devised model contains no gradient terms and is much simpler than those in previous models.

Although previous models were criticized for the inclusion of velocity gradient terms into the source term, it doesn't mean their benefits can be ignored. Since these models are within the framework of the standard LB method, the general philosophy of the LB method is retained. When the distribution function involves the coordinate $r$, people may be bewildered by problems that are seemingly inconsistent with the philosophy of the LB method, such as: why the (radial) coordinate of one node can be propagated to its neighboring node with the particle? In addition, the singularity problem at the axis ($r = 0$) cannot be solved. In numerical applications, this problem is found to cause inconvenience while treating boundary conditions. On the contrary, this problem can be appropriately handled in previous models without affecting the easy treatment of boundary conditions [26-29, 33, 34].

It is generally expected that a more consistent axisymmetric LB scheme can be established if the problems that plague previous schemes are overcome. However, from the available current literature on the axisymmetric LB method, people may conclude that it is impossible to have such an axisymmetric LB scheme. Hence, in this paper, we aim to develop an improved axisymmetric LB scheme based on previous studies and to show that constructing a simple axisymmetric LB scheme within the framework of the standard LB method is possible. The rest of the paper is organized as follows. The macroscopic governing equations for incompressible axisymmetric flows and an original axisymmetric LB scheme are described in Sec. II. The improved scheme is proposed in Sec. III. Without loss of generality, both the BGK and MRT collision models will be considered. In Sec. IV, the numerical validation is presented. Finally, Section V concludes the paper.

## II. MACROSCOPIC EQUATIONS AND AN ORIGINAL AXISYMMETRIC LB SCHEME

The problem of laminar axisymmetric flows of an incompressible fluid with an axis in the $z$ direction is considered. The macroscopic equations for incompressible axisymmetric flows in cylindrical coordinates are given as follows [27, 35, 39]:

$$\partial_j u_j = -u_r/r \tag{1}$$

$$\rho\left[\partial_t u_i + \partial_j\left(u_i u_j\right)\right] = -\partial_i p + \mu \partial_j^2 u_i + \frac{\mu}{r}\partial_r u_i - \frac{\rho u_i u_r}{r} - \frac{\mu u_i}{r^2}\delta_{ir}, \tag{2}$$

where $i$, $j$ indicate the $r$ or $z$ component, $\mu$ is the dynamic viscosity, and $\delta_{ir}$ is the Kronecker delta with two indices. Bearing in mind that, in the standard LB method the recovered macroscopic momentum equation is

$$\partial_t\left(\rho u_i\right) + \partial_j\left(\rho u_i u_j\right) = -\partial_i p + \partial_j\left[\mu\left(\partial_i u_j + \partial_j u_i\right)\right], \tag{3}$$

therefore we need to rewrite Eq. (2) as

$$\rho\left[\partial_t u_i + \partial_j\left(u_i u_j\right)\right] = -\partial_i p + \mu \partial_j\left(\partial_j u_i + \partial_i u_j\right) + \frac{\mu}{r}\left(\partial_r u_i + \partial_i u_r\right) - \frac{\rho u_i u_r}{r} - \frac{2\mu u_i}{r^2}\delta_{ir}. \tag{4}$$

The result $-\mu\partial_j\left(\partial_i u_j\right) = -\mu\partial_i\left(\partial_j u_j\right) = \mu\left(\partial_i u_r - u_i \delta_{ir}/r\right)/r$ has been used in the above derivation.

In Ref. [33], Premnath and Abraham adopted the following evolution equation for axisymmetric flows by integrating the collision and source terms with the trapezium rule:

$$f_\alpha(\boldsymbol{x}+\boldsymbol{e}_\alpha\delta_t, t+\delta_t) - f_\alpha(\boldsymbol{x},t) = \frac{1}{2}\left[\Omega_\alpha\big|_{(\boldsymbol{x},t)} + \Omega_\alpha\big|_{(\boldsymbol{x}+\boldsymbol{e}_\alpha\delta_t, t+\delta_t)}\right]$$
$$+ \frac{\delta_t}{2}\left[G_\alpha\big|_{(\boldsymbol{x},t)} + G_\alpha\big|_{(\boldsymbol{x}+\boldsymbol{e}_\alpha\delta_t, t+\delta_t)}\right], \quad (5)$$

where $\Omega_\alpha = -(f_\alpha - f_\alpha^{eq})/\tau$, $f_\alpha$ is the discrete single-particle density distribution function, $\boldsymbol{x}$ is the spatial vector, i.e., $\boldsymbol{x} = (r,z)$, $\boldsymbol{e}_\alpha = (e_{\alpha r}, e_{\alpha z})$ is the velocity vector of a particle in the $\alpha$ link, $\delta_t$ is the time step, $\tau$ is the dimensionless relaxation time, and $f_\alpha^{eq}$ is the equilibrium distribution, which can be given by

$$f_\alpha^{eq} = w_\alpha \rho \left[1 + \frac{(\boldsymbol{e}_\alpha \cdot \boldsymbol{u})}{c_s^2} + \frac{(\boldsymbol{e}_\alpha \cdot \boldsymbol{u})^2}{2c_s^4} - \frac{u^2}{2c_s^2}\right], \quad (6)$$

for the two-dimensional nine-velocity (D2Q9) lattice [6], where $c_s = c/\sqrt{3}$ ($c = \delta_x/\delta_t$) is the sound speed and the weights $w_\alpha$ are given by $w_0 = 4/9$, $w_{1-4} = 1/9$, and $w_{5-8} = 1/36$. If the axisymmetric contributions of surface tension and phase segregation effects are not considered, the source term $G_\alpha$ is

$$G_\alpha = -w_\alpha \frac{\rho u_r}{r} + \frac{(e_{\alpha i} - u_i)}{\rho c_s^2} f_\alpha^{eq} \left[\frac{\mu}{r}(\partial_r u_i + \partial_i u_r) - \frac{2\mu u_i}{r^2}\delta_{ir} - \frac{\rho u_i u_r}{r}\right]. \quad (7)$$

The Chapman-Enskog analysis [40, 41] of Eq. (5) can be found in Ref. [33]. The implicitness of Eq. (5) is eliminated with $\tilde{f}_\alpha = f_\alpha - 0.5\Omega_\alpha - 0.5\delta_t G_\alpha$ [31]:

$$\tilde{f}_\alpha(\boldsymbol{x}+\boldsymbol{e}_\alpha\delta_t, t+\delta t) - \tilde{f}_\alpha(\boldsymbol{x},t) = -\omega\left[\tilde{f}_\alpha(\boldsymbol{x},t) - f_\alpha^{eq}(\boldsymbol{x},t)\right] + (1-0.5\omega)\delta_t G_\alpha(\boldsymbol{x},t), \quad (8)$$

where $\omega = 1/(\tau + 0.5)$. The macroscopic density and velocities are calculated by

$$\rho = \sum_\alpha \tilde{f}_\alpha - \frac{\delta_t}{2}\frac{\rho u_r}{r}, \quad (9)$$

$$\rho u_i = \sum_\alpha e_{\alpha i}\tilde{f}_\alpha + \frac{\delta_t}{2}\left[\frac{\mu}{r}(\partial_r u_i + \partial_i u_r) - \frac{\rho u_i u_r}{r} - \frac{2\mu u_i}{r^2}\delta_{ir}\right]. \quad (10)$$

## III. IMPROVED LB SCHEME FOR INCOMPRESSIBLE AXISYMMETRIC FLOWS

### A. BGK collision model

In this section, an improved axisymmetric LB scheme will be developed based on the above original scheme. Actually, from Eqs. (4), (7), and (10), it can be seen that, if we want to devise a simple axisymmetric LB scheme based on the standard LB method, the term $\mu(\partial_r u_i + \partial_i u_r)/r$ in the macroscopic axisymmetric momentum equation should be recovered in such a way that the difficulties arising from this term can be avoided. Motivated by our recent work [42], we propose the following evolution equation:

$$f_\alpha(\mathbf{x}+\mathbf{e}_\alpha \delta_t, t+\delta_t) - f_\alpha(\mathbf{x},t) = \frac{1}{2}\left[\Omega_\alpha|_{(\mathbf{x},t)} + \Omega_\alpha|_{(\mathbf{x}+\mathbf{e}_\alpha \delta_t, t+\delta_t)}\right]$$
$$+ \frac{\delta_t}{2}\left[S_\alpha|_{(\mathbf{x},t)} + S_\alpha|_{(\mathbf{x}+\mathbf{e}_\alpha \delta_t, t+\delta_t)}\right] - \delta_t \frac{e_{\alpha r}}{r}(f_\alpha - f_\alpha^{eq}). \quad (11)$$

Note that, in the Chapman-Enskog procedure, the last term on the right-hand side of Eq. (11) will exist in the second-order expansion of the evolution equation. Then no discrete lattice effects need to be considered. The source term $S_\alpha$ is given by

$$S_\alpha = \left[\frac{(e_{\alpha i} - u_i)F_i}{\rho c_s^2} - \frac{u_r}{r}\right]f_\alpha^{eq}, \quad F_i = -\frac{2\mu u_i \delta_{ir}}{r^2}. \quad (12)$$

Here it can be seen that $S_\alpha$ is simple and contains no velocity gradient terms. According to He *et al*. [31], the implicitness of Eq. (11) can be removed with a new distribution function $\hat{f}_\alpha = f_\alpha - 0.5\Omega_\alpha - 0.5\delta_t S_\alpha$, from which the following LB scheme can be obtained:

$$\hat{f}_\alpha(\mathbf{x}+\mathbf{e}_\alpha \delta_t, t+\delta t) - \hat{f}_\alpha(\mathbf{x},t) = -\omega_f\left[\hat{f}_\alpha(\mathbf{x},t) - f_\alpha^{eq}(\mathbf{x},t)\right] + (1-0.5\omega_f)\delta_t S_\alpha(\mathbf{x},t), \quad (13)$$

where $\omega_f = \left[1+(\tau \delta_t e_{\alpha r}/r)\right]/(\tau+0.5)$. The macroscopic variables are defined as

$$\rho = \sum_\alpha \hat{f}_\alpha - \frac{\delta_t}{2}\frac{\rho u_r}{r}, \quad (14)$$

$$\rho u_i = \sum_\alpha e_{\alpha i}\hat{f}_\alpha + \frac{\delta_t}{2}\left[-\frac{\rho u_i u_r}{r} - \frac{2\mu u_i}{r^2}\delta_{ir}\right]. \quad (15)$$

Multiplying Eq. (14) with $u_i$ and then substituting the result into Eq. (15), we can obtain

$$u_i = \frac{\sum_\alpha e_{\alpha i}\hat{f}_\alpha}{\left[\sum_\alpha \hat{f}_\alpha + (\delta_t \mu/r^2)\delta_{ir}\right]}. \quad (16)$$

From Eq. (14), the density is given by

$$\rho = \frac{\sum_\alpha \hat{f}_\alpha}{1+(0.5\delta_t u_r/r)}. \tag{17}$$

In the incompressible limit [43] (i.e., $\rho = \rho_0 + \delta\rho \approx \rho_0$ and $\delta\rho$ is of the order $Ma^2$, where Ma is the Mach number), the viscosity $\mu$ used in $F_i$ and Eq. (16) is replaced with $\mu_0$. In summary, equation (13) together with Eqs. (12), (16), and (17) constitutes an improved axisymmetric LB-BGK scheme.

For the sake of demonstrating that the corresponding macroscopic equations can be correctly recovered in the limit of small Mach number, we proceed to perform the Chapman-Enskog analysis of the evolution equation. First, taking a second-order Taylor series expansion to Eq. (11) in time and space around point $(\mathbf{x},t)$, we have

$$\delta_t(\partial_t + \mathbf{e}_\alpha \cdot \nabla)f_\alpha + \frac{\delta_t^2}{2}(\partial_t + \mathbf{e}_\alpha \cdot \nabla)^2 f_\alpha = -\frac{1}{\tau}(f_\alpha - f_\alpha^{eq}) - \frac{\delta_t}{2\tau}(\partial_t + \mathbf{e}_\alpha \cdot \nabla)(f_\alpha - f_\alpha^{eq})$$
$$+\delta_t S_\alpha + \frac{\delta_t^2}{2}(\partial_t + \mathbf{e}_\alpha \cdot \nabla)S_\alpha - \frac{e_{\alpha r}}{r}\delta_t(f_\alpha - f_\alpha^{eq}) + O(\delta_t^3), \tag{18}$$

where $\nabla = (\partial_r, \partial_z)$ is the spatial gradient operator. According to the Chapman-Enskog expansion [25, 40, 41], the time derivative, the distribution function, and the source term can be written as

$$\partial_t = \partial_{t0} + \delta_t \partial_{t1}, \quad f_\alpha = f_\alpha^{(0)} + \delta_t f_\alpha^{(1)} + \delta_t^2 f_\alpha^{(2)}, \quad S_\alpha = S_\alpha^{(0)} + \delta_t S_\alpha^{(1)}, \tag{19}$$

where $S_\alpha^{(0)} = -u_r f_\alpha^{eq}/r$ and $S_\alpha^{(1)} = \frac{(e_{\alpha i}-u_i)f_\alpha^{eq}}{\rho c_s^2}\left(-\frac{2\rho c_s^2 \tau u_i \delta_{ir}}{r^2}\right)$. With these multi-scale expansions, we can rewrite Eq. (18) in the consecutive orders of $\delta_t$:

$$O(\delta_t^0): f_\alpha^{(0)} = f_\alpha^{eq}, \tag{20}$$

$$O(\delta_t^1): (\partial_{t0} + \mathbf{e}_\alpha \cdot \nabla)f_\alpha^{(0)} + \frac{1}{\tau}f_\alpha^{(1)} = S_\alpha^{(0)}, \tag{21}$$

$$O(\delta_t^2): \partial_{t1}f_\alpha^{(0)} + (\partial_{t0} + \mathbf{e}_\alpha \cdot \nabla)f_\alpha^{(1)} + \frac{1}{2}(\partial_{t0} + \mathbf{e}_\alpha \cdot \nabla)^2 f_\alpha^{(0)} + \frac{1}{\tau}f_\alpha^{(2)} + \frac{1}{2\tau}(\partial_{t0} + \mathbf{e}_\alpha \cdot \nabla)f_\alpha^{(1)}$$
$$= \frac{1}{2}(\partial_{t0} + \mathbf{e}_\alpha \cdot \nabla)S_\alpha^{(0)} + S_\alpha^{(1)} - \frac{e_{\alpha r}}{r}f_\alpha^{(1)}. \tag{22}$$

Using Eq.(21), Eq. (22) can be rewritten as

$$\partial_{t1} f_\alpha^{(0)} + \left(\partial_{t0} + \boldsymbol{e}_\alpha \cdot \nabla\right) f_\alpha^{(1)} + \frac{1}{\tau} f_\alpha^{(2)} = -\frac{e_{\alpha r}}{r} f_\alpha^{(1)} + S_\alpha^{(1)}. \tag{23}$$

Summations of Eq. (21) and Eq. (23) lead to, respectively

$$\partial_{t0} \rho + \partial_j \left(\rho u_j\right) = -\frac{\rho u_r}{r}, \tag{24}$$

$$\partial_{t1} \rho = 0. \tag{25}$$

Combining the above two equations ($\partial_t = \partial_{t0} + \delta_t \partial_{t1}$) gives

$$\partial_t \rho + \partial_j \left(\rho u_j\right) = -\frac{\rho u_r}{r}. \tag{26}$$

Taking the first-order moment, $\sum_\alpha e_{\alpha i}(\cdot)$, of Eqs. (21) and (23), respectively, we get

$$\partial_{t0}\left(\rho u_i\right) + \partial_j\left(\rho u_i u_j\right) = -\partial_i p - \frac{\rho u_i u_r}{r}, \tag{27}$$

$$\partial_{t1}\left(\rho u_i\right) + \partial_j\left(\sum_\alpha e_{\alpha i} e_{\alpha j} f_\alpha^{(1)}\right) = -\frac{1}{r}\sum_\alpha e_{\alpha i} e_{\alpha r} f_\alpha^{(1)} - \frac{2\rho c_s^2 \tau u_i}{r^2}\delta_{ir}. \tag{28}$$

From Eq. (21), it is obtained that

$$\sum_\alpha e_{\alpha i} e_{\alpha j} f_\alpha^{(1)} = -\tau\left(\partial_{t0} \Pi_{ij}^{(0)} + \partial_k P_{ijk}^{(0)} - \sum_\alpha e_{\alpha i} e_{\alpha j} S_\alpha^{(0)}\right), \tag{29}$$

where $\Pi_{ij}^{(0)} = \sum_\alpha e_{\alpha i} e_{\alpha j} f_\alpha^{(0)}$ and $P_{ijk}^{(0)} = \sum_\alpha e_{\alpha i} e_{\alpha j} e_{\alpha k} f_\alpha^{(0)}$. For the D2Q9 lattice model, $\Pi_{ij}^{(0)}$ and $P_{ijk}^{(0)}$ are given by

$$\Pi_{ij}^{(0)} = \rho u_i u_j + \rho c_s^2 \delta_{ij}, \quad P_{ijk}^{(0)} = \rho c_s^2 \left(u_i \delta_{jk} + u_j \delta_{ik} + u_k \delta_{ij}\right). \tag{30}$$

Some standard algebra will show that

$$\partial_{t0} \Pi_{ij}^{(0)} = c_s^2 \delta_{ij} \partial_{t0} \rho - c_s^2 u_j \partial_i \rho - c_s^2 u_i \partial_j \rho + O\left(u^3\right), \tag{31}$$

$$\partial_k P_{ijk}^{(0)} = c_s^2 \delta_{ij} \partial_k \left(\rho u_k\right) + \rho c_s^2 \partial_j u_i + \rho c_s^2 \partial_i u_j + c_s^2 u_i \partial_j \rho + c_s^2 u_j \partial_i \rho. \tag{32}$$

With the above results, we have

$$\partial_{t0} \Pi_{ij}^{(0)} + \partial_k P_{ijk}^{(0)} = c_s^2 \delta_{ij}\left[\partial_{t0}\rho + \partial_k\left(\rho u_k\right)\right] + \rho c_s^2 \left(\partial_j u_i + \partial_i u_j\right). \tag{33}$$

Using Eq. (33) with Eq. (24) and noting that $\sum_\alpha e_{\alpha i} e_{\alpha j} S_\alpha^{(0)} = -c_s^2 \delta_{ij}\, \rho u_r/r + O\left(u^3\right)$, we can simplify Eq. (29) to

$$\sum_\alpha e_{\alpha i} e_{\alpha j} f_\alpha^{(1)} = -\tau \rho c_s^2 \left(\partial_j u_i + \partial_i u_j\right). \tag{34}$$

Substituting Eq. (34) into Eq. (28) yields

$$\partial_{t1}(\rho u_i) = \partial_j \left[ \tau \rho c_s^2 \left( \partial_j u_i + \partial_i u_j \right) \right] + \frac{1}{r} \tau \rho c_s^2 \left( \partial_r u_i + \partial_i u_r \right) - \frac{2\rho c_s^2 \tau u_i}{r^2} \delta_{ir}. \qquad (35)$$

Combining Eq. (35) with Eq. (27) ($\partial_t = \partial_{t0} + \delta_t \partial_{t1}$), we can obtain

$$\partial_t (\rho u_i) + \partial_j (\rho u_i u_j) = -\partial_i p + \partial_j \left[ \mu \left( \partial_j u_i + \partial_i u_j \right) \right] + \frac{\mu}{r} \left( \partial_r u_i + \partial_i u_r \right) - \frac{\rho u_i u_r}{r} - \frac{2\mu u_i}{r^2} \delta_{ir}, \qquad (36)$$

where $\mu = \tau \rho c_s^2 \delta_t$. Clearly, in the incompressible limit ($\rho \approx \rho_0$), Eqs. (26) and (36) reduce to the axisymmetric continuity equation (1) and the momentum equation (4), respectively.

Now a brief comparison between the improved and original schemes is made. First, both schemes are within the framework of the standard LB method using the single-particle density distribution function and have a simple structure so that the general benefits of the standard LB method are retained. On the other hand, in the improved scheme, the term $\mu(\partial_r u_i + \partial_i u_r)/r$ is recovered in an efficient way that is consistent with the philosophy of the LB method. As a consequence, the source term and the calculations of macroscopic variables are greatly simplified. Accordingly, the problems that plague the original scheme are overcome.

### B. MRT collision model

#### 1. MRT-LB method

In Ref. [34], the MRT collision model, which is an important extension of the relaxation LB method proposed by Higuera [2, 3], has been employed to construct an axisymmetric MRT-LB scheme based on the above-mentioned original scheme. Much research has shown that the MRT collision model can significantly improve the numerical stability of LB schemes by carefully separating the relaxation times of hydrodynamic and non-hydrodynamic moments. A detailed description of the MRT-LB method can be found in Refs. [8-13]. According to Refs. [12, 13, 34], a D2Q9 MRT-LB scheme with a semi-implicit treatment of the source term is given by

$$f_\alpha(\bm{x}+\bm{e}_\alpha\delta_t,t+\delta_t)-f_\alpha(\bm{x},t)=-\bar{\Lambda}_{\alpha\beta}\left(f_\beta-f_\beta^{eq}\right)\Big|_{(\bm{x},t)}+\frac{\delta_t}{2}\left[S_\alpha\big|_{(\bm{x},t)}+S_\alpha\big|_{(\bm{x}+\bm{e}_\alpha\delta_t,t+\delta_t)}\right], \tag{37}$$

where $\bar{\Lambda}=\mathbf{M}^{-1}\Lambda\mathbf{M}$ is the collision matrix, in which $\Lambda=\mathrm{diag}(s_\rho,s_e,s_\varepsilon,s_j,s_q,s_j,s_q,s_v,s_v)$ is a diagonal Matrix and $\mathbf{M}$ is a orthogonal transformation matrix (see Ref. [9]).

Through the transformation matrix, the distribution function $f_\alpha$ and its equilibrium distribution $f_\alpha^{eq}$ can be projected onto the moment space with $\mathbf{m}=\mathbf{Mf}$ and $\mathbf{m}^{eq}=\mathbf{Mf}^{eq}$, where $\mathbf{f}=(f_0,f_1,\cdots,f_8)^{\mathrm{T}}$ and $\mathbf{f}^{eq}=(f_0^{eq},\cdots,f_8^{eq})^{\mathrm{T}}$. For the D2Q9 lattice model, $\mathbf{m}$ and $\mathbf{m}^{eq}$ are given by

$$\mathbf{m}=(\rho,e,\varepsilon,j_x,q_x,j_y,q_y,p_{xx},p_{xy})^{\mathrm{T}}, \tag{38}$$

$$\begin{aligned}\mathbf{m}^{eq}&=\left(\rho,e^{eq},\varepsilon^{eq},j_x,q_x^{eq},j_y,q_y^{eq},p_{xx}^{eq},p_{xy}^{eq}\right)^{\mathrm{T}}\\&=\rho\left(1,-2+3u^2,1-3u^2,u_x,-u_x,u_y,-u_y,u_x^2-u_y^2,u_xu_y\right)^{\mathrm{T}},\end{aligned} \tag{39}$$

where $\rho$ is the density; $e$ is the energy mode; $\varepsilon$ is related to energy square; $(j_x,j_y)$ are the momentum components; $(q_x,q_y)$ correspond to energy flux; and $(p_{xx},p_{xy})$ are related to the diagonal and off-diagonal components of the stress tensors [9].

Because of the implicit treatment of the source term, Eq. (37) cannot be directly applied in numerical simulations. The following explicit MRT-LB scheme can be obtained with $\bar{f}_\alpha=f_\alpha-0.5\delta_t S_\alpha$ [12, 13]:

$$\bar{f}_\alpha(\bm{x}+\bm{e}_\alpha\delta_t,t+\delta_t)=\bar{f}_\alpha(\bm{x},t)-\bar{\Lambda}_{\alpha\beta}\left(\bar{f}_\beta-f_\beta^{eq}\right)\Big|_{(\bm{x},t)}+\delta_t\left(S_\alpha-0.5\bar{\Lambda}_{\alpha\beta}S_\beta\right)\Big|_{(\bm{x},t)}. \tag{40}$$

Usually, as shown in Ref. [9], the collision process of MRT-LB schemes is carried out in the moment space

$$\bar{\mathbf{m}}^+=\bar{\mathbf{m}}-\Lambda\left(\bar{\mathbf{m}}-\mathbf{m}^{eq}\right)+\delta_t\left(\mathbf{I}-\frac{\Lambda}{2}\right)\tilde{\mathbf{S}}, \tag{41}$$

where $\bar{\mathbf{m}}=\mathbf{M}\bar{\mathbf{f}}$ and $\tilde{\mathbf{S}}=\mathbf{MS}$, in which $\mathbf{S}=(S_0,S_1,\cdots,S_8)^{\mathrm{T}}$, while the streaming process is implemented in the velocity space

$$\bar{f}_\alpha(\bm{x}+\bm{e}_\alpha\delta_t,t+\delta_t)=\bar{f}_\alpha^+(\bm{x},t), \tag{42}$$

where $\bar{\mathbf{f}}^+ = \mathbf{M}^{-1}\bar{\mathbf{m}}^+$. According to Eq. (41), the collision process of the momentum components can be written as

$$\bar{j}_x^+ = \bar{j}_x - s_j\left(\bar{j}_x - j_x^{eq}\right) + \delta_t\left(1 - 0.5 s_j\right)\tilde{S}_3, \tag{43}$$

$$\bar{j}_y^+ = \bar{j}_y - s_j\left(\bar{j}_y - j_y^{eq}\right) + \delta_t\left(1 - 0.5 s_j\right)\tilde{S}_5. \tag{44}$$

The macroscopic equations recovered form MRT-LB schemes can also be derived through the Chapman-Enskog analysis, which can be implemented in the moment space. For the details of this procedure, readers are referred to Refs. [9, 12, 13, 44]. Several relationships are given below considering that they will be used in the next subsection:

$$-s_e e^{(1)} = 2\rho\left(\partial_x u_x + \partial_y u_y\right),\quad -s_v p_{xx}^{(1)} = \frac{2\rho}{3}\left(\partial_x u_x - \partial_y u_y\right),\quad -s_v p_{xy}^{(1)} = \frac{\rho}{3}\left(\partial_x u_y + \partial_y u_x\right), \tag{45}$$

where $e^{(1)}$, $p_{xx}^{(1)}$, and $p_{xy}^{(1)}$ are defined as $\delta_t \mathbf{m}^{(1)} \approx \mathbf{m} - \mathbf{m}^{eq}$, corresponding to $\delta_t f_\alpha^{(1)} \approx f_\alpha - f_\alpha^{eq}$. Note that terms of $O(\mathrm{Ma}^3)$ have been neglected in Eq. (45).

*2. Axisymmetric MRT-LB scheme*

It is known that, with the benefit of using MRT collision model, the collision process of each moment can be manipulated independently in the moment space. The approach of modifying the collision process to adjust macroscopic equations has been used in Ref. [45], in which the two moments related to the energy flux were modified to achieve a consistent viscosity in the macroscopic momentum and energy equations.

In the present work, it is found that the collision process of the momentum components can be appropriately manipulated to recover the velocity gradient term $\mu(\partial_r u_i + \partial_i u_r)/r$ in the axisymmetric momentum equation. To this end, we need to evaluate $\mu(\partial_r u_i + \partial_i u_r)/r$ in a way consistent with the philosophy of the MRT-LB method. Note that, from Eq. (45), we have

$$-\left(\frac{1}{6}s_e e^{(1)} + \frac{1}{2}s_v p_{xx}^{(1)}\right) = \frac{\rho}{3}\left(\partial_x u_x + \partial_x u_x\right). \tag{46}$$

Using Eqs. (45) and (46), we can obtain

$$\frac{\mu}{r}\left(\partial_r u_r + \partial_r u_r\right) = -\frac{\gamma}{s_v}\left(\frac{1}{6}s_e e^{(1)} + \frac{1}{2}s_v p_{xx}^{(1)}\right), \tag{47}$$

$$\frac{\mu}{r}\left(\partial_r u_z + \partial_z u_r\right) = -\gamma p_{xy}^{(1)}, \tag{48}$$

where $\mu = (1/s_v - 0.5)\delta_t \rho/3$, $\gamma = (1-0.5 s_v)\delta_t/r$, and $(r, z)$ correspond to $(x, y)$. Thus we can modify the collision process of the momentum components as follows:

$$\left(\bar{j}_x^+\right)_{new} = \bar{j}_x^+ - \delta_t \gamma\left[\left(s_e/6s_v\right)e^{(1)} + 0.5 p_{xx}^{(1)}\right], \tag{49}$$

$$\left(\bar{j}_y^+\right)_{new} = \bar{j}_y^+ - \delta_t \gamma p_{xy}^{(1)}. \tag{50}$$

Here $e^{(1)}$, $p_{xx}^{(1)}$, and $p_{xy}^{(1)}$ are given by $\mathbf{m}^{(1)} = \bar{\mathbf{m}}^{(1)} + 0.5\tilde{\mathbf{S}}$ (recalling the equation $\bar{f}_\alpha = f_\alpha - 0.5\delta_t S_\alpha$), in which $\bar{\mathbf{m}}^{(1)}$ is approximated by $\delta_t \bar{\mathbf{m}}^{(1)} \approx \bar{\mathbf{m}} - \mathbf{m}^{eq}$. The source term in the moment space is $\tilde{\mathbf{S}} = (u_r/r)\mathbf{m}^{eq} + \tilde{\mathbf{S}}'$ with $\tilde{S}'_\alpha$ given as follows [12]:

$$\tilde{S}'_0 = 0, \quad \tilde{S}'_1 = 6\mathbf{u}\cdot\mathbf{F}, \quad \tilde{S}'_2 = -6\mathbf{u}\cdot\mathbf{F},$$

$$\tilde{S}'_3 = F_x, \quad \tilde{S}'_4 = -F_x, \quad \tilde{S}'_5 = F_y, \quad \tilde{S}'_6 = -F_y,$$

$$\tilde{S}'_7 = 2\left(u_x F_x - u_y F_y\right), \quad \tilde{S}'_8 = \left(u_x F_y + u_y F_x\right), \tag{51}$$

where $F_x = F_r$ and $F_y = F_z$ are given in Eq. (12). It can be readily proved that, with such a choice of the source term, the relationships shown in Eq. (45) will not change. Finally, equations (41) and (42) together with the modified collision process, Eqs. (49) and (50), constitute a consistent axisymmetric MRT-LB scheme for incompressible axisymmetric flows. The macroscopic variables are calculated by Eqs. (16) and (17) through replacing $\hat{f}_\alpha$ with $\bar{f}_\alpha$.

### C. Extension to axisymmetric rotational flows

By including the effect of azimuthal rotation, the proposed scheme can be applied to axisymmetric rotational flows. The macroscopic governing equation for azimuthal velocity $u_\theta$ in cylindrical coordinates is given by [39]

$$\rho\left[\partial_t u_\theta + \partial_j\left(u_j u_\theta\right)\right] = \mu \partial_j\left(\partial_j u_\theta\right) + \frac{\mu}{r}\partial_r u_\theta - \frac{2\rho u_\theta u_r}{r} - \frac{\mu u_\theta}{r^2}. \quad (52)$$

It is seen that the above equation is an advection-diffusion equation. Usually, a D2Q4 or D2Q5 lattice model is enough for an advection-diffusion equation in terms of the computational accuracy as well as the computational efficiency [46]. Furthermore, as suggested in Ref. [36], the source term of an advection-diffusion equation can be treated more simply than the usual forcing strategy. Hence in this study the following evolution equation with a D2Q4 lattice ($e_\alpha : \alpha = 1, 2, 3, 4.$) is adopted to solve the azimuthal velocity:

$$g_\alpha\left(\boldsymbol{x} + \boldsymbol{e}_\alpha \delta_t, t + \delta_t\right) - g_\alpha\left(\boldsymbol{x}, t\right) = -\omega_g\left[g_\alpha\left(\boldsymbol{x}, t\right) - g_\alpha^{eq}\left(\boldsymbol{x}, t\right)\right] + \delta_t S_\alpha^g\left(\boldsymbol{x}, t\right). \quad (53)$$

Here $g_\alpha$ is the distribution function for azimuthal velocity, $\omega_g = \left[1 + \left(\tau_g \delta_t e_{\alpha r}/r\right)\right]/\left(\tau_g + 0.5\right)$, and

$$g_\alpha^{eq} = \frac{\rho u_\theta}{4}\left[1 + 2\frac{\left(\boldsymbol{e}_\alpha \cdot \boldsymbol{u}\right)}{c^2}\right], \quad S_\alpha^g = -\frac{1}{r}\left(2u_r + \frac{\nu}{r}\right)g_\alpha^{eq}, \quad (54)$$

where $\nu = \delta_t \tau_g c^2 / 2$ is the kinematic viscosity and $\tau_g = 2\tau/3$. Clearly, the source term $S_\alpha^g$ is also simple and contains no gradient terms. Note that, when the effect of azimuthal rotation is considered, an "inertial force" $\rho u_\theta^2 \delta_{ir}/r$ should be included in $F_i$ of Eq. (12). Moreover, in the incompressible limit, the density $\rho$ in the equilibrium distribution function $g_\alpha^{eq}$ can be directly replaced by $\rho_0$, and then the macroscopic azimuthal velocity is calculated by $u_\theta = \sum_\alpha g_\alpha / \rho_0$.

## IV. NUMERICAL VALIDATION

### A. Hagen-Poiseuille flow and 3D Womersley flow

To validate the proposed scheme, numerical simulations are carried out for some typical axisymmetric flows. First, we consider the Hagen-Poiseuille flow, which is an axisymmetric steady, laminar flow of a viscous fluid through a pipe of uniform circular cross-section and driven by a constant external force in the axial direction. The analytical solution for the axial velocity of the Hagen-Poiseuille flow is given by

$$u_z(r) = U_0\left(1 - \frac{r^2}{R^2}\right), \tag{55}$$

where $U_0 = aR^2/(4\mu)$ is the maximum axial velocity in the pipe, $a$ is the external force, and $R$ is the radius of the pipe.

In the simulation, we adopt a $N_z \times N_r = 40 \times 20$ lattice ($N_z$ and $N_r$ exclude the extra layers outside the boundaries) with a line of symmetry at $r = 0$ and a solid wall at $r = 20$. The no-slip boundary condition is imposed along the solid wall [47], the periodic boundary conditions are applied to the inlet and outlet, and the specular reflection boundary [26] is employed along the axisymmetric line. The singularity at $r = 0$ is treated following previous studies. In Refs. [29, 33], the source term at $r = 0$ was evaluated with the L'Hôpital's rule; after applying this rule, the source term was set to be zero. Similarly, in Ref. [26], all the terms related to $(1/r)$ were only applied at the position of $r \neq 0$. In other words, these terms were approximately taken as zero at $r = 0$. In the present paper, a similar treatment is adopted. The maximum velocity $U_0$ is set to be 0.05 with $a = 10^{-4}$ and $\mu = 0.2$. The numerical axial velocity is shown in Fig. 1, in which the analytical solution is also presented for comparison. It is observed that the numerical result agrees well with the analytical one.

When the constant force in the Hagen-Poiseuille flow oscillates with a period $T$, the flow will become a 3D Womersley flow, which is an unsteady axisymmetric flow in a circular pipe driven by a periodic force $a = a_0 \cos(\omega t)$, where $a_0$ is the maximum amplitude and the $\omega = 2\pi/T$ is the angular frequency [48]. The Reynolds number is defined as $\text{Re} = \rho_0 U_c D/\mu$ with the characteristic length $D = 2R$ and the characteristic velocity $U_c = a_0 \alpha^2/(4\rho_0 \omega)$. Here $\alpha = R\sqrt{\rho_0 \omega/\mu}$ is the

Womersley number. The analytical solution for the Womersley flow is [27, 48]

$$u_z(r,t) = \text{Re}\left\{\frac{a_0}{i\rho_0\omega}\left[1 - \frac{J_0(r\phi/R)}{J_0(\phi)}\right]e^{i\omega t}\right\}, \quad (56)$$

where $J_0$ is the zeroth order Bessel function of the first type and $i$ is the imaginary unit, $\phi = (-\alpha + i\alpha)/\sqrt{2}$. Note that Re in Eq. (56) denotes the real part of a complex number rather than the Reynolds number.

In this test, the boundary conditions and the grid system are the same as those used in the simulation of Hagen-Poiseuille flow. Following previous studies, in the computation we set $\text{Re} = 1200$, $T = 1200$, $a_0 = 10^{-3}/3$, $\alpha = 7.927$, $\rho = \rho_0 = 1$, and $\mu = 0.1/3$; the simulations begin with an initial condition of zero velocity and the numerical results at different times are obtained after running ten periods. Figures 2 and 3 compare the axial velocity predicted by the present scheme with the analytical solutions. It can be found that the numerical results are in excellent agreement with the analytical one. To quantify the results, a relative error defined as $\xi = \sum_i |u(r_i) - u_a(r_i)| / \sum_i |u_a(r_i)|$ (the subscript $a$ denotes "analytical") is used to compare the present solution with the numerical solution shown in Ref. [27], which is a better one among the results reported in previous studies. The global error $\langle \xi \rangle$ is the average of $\xi$ over the period. The present $\langle \xi \rangle$ is 0.33%, which is smaller than the result 1.3% reported in Ref. [27].

**B. Axisymmetric rotational flows**

In this subsection, the Wheeler benchmark problem in the Czochralski crystal growth [26, 49-51] and the lid-driven rotational flow in cylindrical cavities [52-58] are taken as the test examples to validate the capability of the proposed scheme for the simulation of axisymmetric rotational flows. The configuration of the Wheeler problem is descried in Fig. 4. In the problem, a vertical cylindrical crucible of radius $R_c$ filled with a melt to a height $H = R_c$ rotates with an angular velocity $\Omega_c$. On the top of the melt, it is bounded by a coaxial crystal with radius $R_x = \beta R_c$ ($\beta = 0.4$), which rotates with an angular velocity $\Omega_x$. The flow structure depends on the Reynolds number $\text{Re}_c = R_c^2 \Omega_c / \nu$ and $\text{Re}_x = R_c^2 \Omega_x / \nu$.

In the computations, a $N_r \times N_z = 100 \times 100$ lattice is adopted and the value of the characteristic velocity $U_t = \beta R_c \Omega_x$ is taken as $U_t \leq 0.1$ so that the Mach number of the flow is sufficiently small. The relaxation times can be determined with $U_t$ and $\text{Re}_x$. The non-equilibrium extrapolation scheme [59] is employed to treat different boundary conditions of $\hat{f}_\alpha$ and $g_\alpha$, except that the specular reflection boundary is applied to $\hat{f}_\alpha$ along the axisymmetric line. The zero velocities are initialized everywhere. A steady state can be reached after a number of iterations and the convergence criterion is

$$\max \left| \sqrt{\left(u_z^2 + u_r^2\right)^{n+1}} - \sqrt{\left(u_z^2 + u_r^2\right)^n} \right| \leq \varsigma, \tag{57}$$

where $n$ and $n+1$ represent the old and new time levels, respectively. $\varsigma$ is set to be $10^{-8}$ in this test [26].

The streamlines of $(\text{Re}_c, \text{Re}_x) = (10^2, -25)$ and $(10^3, -250)$ are presented in Fig. 5, from which we can see that two vortices with opposite directions appear in the upper left corner and the lower right corner. With the increase of the Reynolds number, the upper left vortex moves towards right corner and the lower right primary vortex moves to left and dominates the whole flow field. These behaviors are also found in previous numerical studies. To quantify the results, the stream function defined as $\partial_r \psi = -r u_z$, $\partial_z \psi = r u_r$ is calculated and Table 1 shows the comparisons of $\psi_{\min}$ and $\psi_{\max}$ between the present results and the results reported in Refs. [26, 50]. Good agreement can be concluded from the table.

In the Wheeler problem, if we set $\text{Re}_c = \Omega_c = 0$ and $R_x = R_c$, the flow will become the lid-driven rotational flow in cylindrical cavities, which is an important generic problem investigated both experimentally [52-54] and numerically [55-58]. The cylindrical cavity rotational flow is known to depend on two parameters, the aspect ratio $A = H/R$ and the Reynolds number $\text{Re} = R^2\Omega/\nu$. In the literature, it has been confirmed that, at certain combinations of $A$ and $\text{Re}$, a recirculation region will form along the axis of the cylinder. Such a recirculation region is called the vortex breakdown bubble. In this study, the cases $\text{Re} = 990$ and $1290$ with $A = 1.5$ are considered following Ref. [58]. The experimental results of these two cases are available [54]. In our simulations, a grid system $N_r \times N_z = 100 \times 150$ is adopted. The boundary treatments are similar to those used in the Wheeler problem. The relaxation time $\tau$ is chosen as $\tau = 0.02$ to ensure the characteristic velocity $U_t = \Omega R \leq 0.1$. The obtained streamlines are presented in Fig. 6. From the figure it is seen that a single vortex breakdown appears at $\text{Re} = 1290$, whereas the result of $\text{Re} = 990$ do not reveal any vortex breakdown. Table 2 shows the magnitude ($u_{z,\max}$) and the location ($h_{\max}/H$) of the maximum axial velocity on the axis, in which the experiment results [54] and the numerical results obtained from 3D-LB model [58] are also listed for comparison. To sum up, the present results are well consistent with the previous ones.

The comparison between the BGK and MRT collision models is conducted through simulating cylindrical cavity rotational flow at $\text{Re} = 1290$ with a low viscosity $\nu = 1.67 \times 10^{-3}$, which corresponds to $\tau = 0.005$ and $s_\nu = 1/(0.5 + 0.005)$. The convergence criterion $\varsigma$ is set to be $10^{-10}$ as the characteristic velocity $U_t$ is greatly decreased. The simulation results are presented in Fig. 7. It can be observed that the BGK model is numerically unstable when $\tau = 0.005$, but the MRT model can give a stable and correct solution under the same condition. The comparison illustrates that the enhanced numerical stability of the MRT model compared with the BGK model in that the MRT model is able to achieve stable results at lower viscosities [12].

Finally, an important issue should be pointed out. Strictly speaking, in the axisymmetric LB-MRT method the bulk viscosity should be equal to the dynamic viscosity, which requires $s_e = s_v$. This is because, for an axisymmetric LB scheme in cylindrical coordinates, the pseudo divergence of velocity $\partial_i u_i = -u_r/r$ is nonzero. While in the Cartesian coordinate system, the real divergence of velocity is zero for incompressible flows. In simulations, an appropriate difference between $s_v$ and $s_e$ may be permitted if needed. However, when $s_e$ significantly differs from $s_v$, large errors will be introduced. The stable solution presented in Fig. 7 is obtained based on $s_e = s_v$. In fact, a MRT collision model with $s_e = s_v$ is similar to a two-relaxation-time (TRT) collision model [14-16], which is an important and natural simplification of MRT collision model [60]. In TRT models, the moments of even ($s_e = s_\varepsilon = s_v$) and odd orders are relaxed at different rates.

## IV. CONCLUSIONS

In this paper, an improved axisymmetric LB scheme has been developed for incompressible axisymmetric flows. It has been shown that constructing a simple axisymmetric LB scheme within the framework of the standard LB method using the single-particle density distribution function is possible. The main strategy is to recover the term $\mu(\partial_r u_i + \partial_i u_r)/r$ in the macroscopic momentum equation in an efficient way that is consistent with the philosophy of the LB method. The Chapman-Enskog analysis has been employed to demonstrate that the macroscopic equations can be correctly recovered in the limit of small Mach number. As a result of the change, the source term becomes simple and contains no velocity gradient terms. Furthermore, the calculations of macroscopic variables are simplified. The singularity problem at the axis is appropriately treated following previous studies, retaining the easy treatment of boundary conditions. In the proposed scheme, both the BGK and MRT collision models have been considered. In addition, its extension to axisymmetric rotational flows is also presented. Numerical simulations have been carried out for some typical axisymmetric flows. The numerical experiments show that the results predicted by the present scheme are in good agreement with the analytical solutions and the results reported in previous studies. The comparison between the BGK and MRT collision models has also been made. It is shown that the MRT collision model exhibits an excellent numerical stability compared with the BGK model when the viscosity approaches zero. This feature makes MRT-LB schemes more useful in practical applications.

## ACKNOWLEDGMENTS

This work was supported by the Key Project of National Natural Science Foundation of China (No.50736005) and the National Basic Research Program of China (973 Program) (No. 2007CB206902).

## References

[1] U. Frisch, B. Hasslacher, and Y. Pomeau, Phys. Rev. Lett., **56**, 1505 (1986).

[2] F. Higuera and J. Jiménez, Europhys. Lett. **9**, 663 (1989).

[3] F. Higuera, S. Succi, and R. Benzi, Europhys. Lett. **9**, 345 (1989).

[4] J. M. V. A. Koelman, Europhys. Lett., **15**, 603 (1991).

[5] S. Chen, H. Chen, D. Martínez, and W. Matthaeus, Phys. Rev. Lett. **67**, 3776 (1991).

[6] Y. H. Qian, D. d'Humières, and P. Lallemand, Europhys. Lett. **17**, 479 (1992).

[7] X. He and L.-S. Luo, Phys. Rev. E **56**, 6811 (1997).

[8] D. d'Humières, in *Rarefied Gas Dynamics: Theory and Simulations*, Prog. Astronaut. Aeronaut. Vol. 159, edited by B. D. Shizgal and D. P. Weaver (AIAA, Washington, D.C., 1992).

[9] P. Lallemand and L.-S. Luo, Phys. Rev. E **61**, 6546 (2000).

[10] D. d'Humières, I. Ginzburg, M. Krafczyk, P. Lallemand, and L.-S. Luo, Phil. Trans. R. Soc. Lond. A **360**, 437 (2002).

[11] P. J. Dellar, J. Comput. Phys. **190**, 351 (2003).

[12] M. E. MaCracken and J. Abraham, Phys. Rev. E **71**, 036701 (2005).

[13] K. N. Premnath and J. Abraham, J. Comput. Phys. **224**, 539 (2007).

[14] I. Ginzburg, Adv. Water Resour. **28**, 1171 (2005).

[15] I. Ginzburg, Adv. Water Resour. **28**, 1196 (2005).

[16] I. Ginzburg, Comm. Comp. Phys. **3**, 427 (2008).

[17] S. Succi, Eur. Phys. J. B **64**, 471 (2008).

[18] R. Benzi, S. Succi, and M. Vergassola, Phys. Rep. **222**, 145 (1992).

[19] S. Chen and G. D. Doolen, Annu. Rev. Fluid Mech., **30**, 329 (1998).

[20] A J. C. Ladd and R. Verberg, J. Stat. Phys. 104, 1191 (2001).

[21] D. Yu, R. Mei, L.-S. Luo, and W. Shyy, Prog. Aerosp. Sci. **39**, 329 (2003).

[22] B. Dünweg and A. J. C. Ladd, Adv. Polym. Sci. 221, 89 (2009).

[23] S. Succi, *The Lattice Boltzmann Equation for Fluid Dynamics and Beyond* (Clarendon Press, Oxford, 2001).

[24] M. C. Sukop and D. T. Jr Thorne, *Lattice Boltzmann modeling: An introduction for geoscientists and engineers* (Springer, Berlin, 2005).

[25] I. Halliday, L. A. Hammond, C. M. Care, K. Good, and A. Stevens, Phys. Rev. E **64**, 011208

(2001).

[26] Y. Peng, C. Shu, Y. T. Chew, and J. Qiu, J. Comput. Phys. **186**, 295 (2003).

[27] T. S. Lee, H. Huang, and C. Shu, Int. J. Mod. Phys. C **17**, 645 (2006).

[28] T. Reis and T. N. Phillips, Phys. Rev. E **75**, 056703 (2007).

[29] T. Reis and T. N. Phillips, Phys. Rev. E **77**, 026703 (2008).

[30] X. He, X. Shan, and G. D. Doolen, Phys. Rev. E **57**, R13 (1998).

[31] X. He, S. Chen, and G. D. Doolen, J. Comput. Phys. **146**, 282 (1998).

[32] Z. Guo, C. Zheng, and B. Shi, Phys. Rev. E **65**, 046308 (2002).

[33] K. N. Premnath and J. Abraham, Phys. Rev. E **71**, 056706 (2005).

[34] S. Mukherjee and J. Abraham, Phys. Rev. E **75**, 026701 (2007).

[35] J. G. Zhou, Phys. Rev. E **78**, 036701 (2008).

[36] S. Chen, J. Tölke, S. Geller, and M. Krafczyk, Phys. Rev. E **78**, 046703 (2008).

[37] S. Chen, J. Tölke, and M. Krafczyk, Phys. Rev. E **79**, 016704 (2009).

[38] Z. Guo, H. Han, B. Shi, and C. Zheng, Phys. Rev. E **79**, 046708 (2009).

[39] F. M. White, *Fluid Mechanics* (5th ed., McGraw-Hill, New York, 2003).

[40] S. Chapman and T. G. Cowling, *The Mathematical Theory of Non-Uniform Gases*, 3rd ed. (Cambridge University Press, Cambridge, UK, 1970).

[41] S. Hou, Q. Zou, S. Chen, G. Doolen, and A. C. Cogley, J. Comput. Phys. **118**, 329 (1995).

[42] Q. Li, Y. L. He, G. H. Tang, and W. Q. Tao, Phys. Rev. E **80**, 037702 (2009).

[43] X. He and L.-S. Luo, J. Stat. Phys. **88**, 927 (1997).

[44] R. Du, B. Shi, and X. Chen, Phys. Lett. A **359**, 564 (2006).

[45] L. Zheng, B. Shi, and Z. Guo, Phys. Rev. E **78**, 026705 (2009).

[46] S. Chen et al., Appl. Math. Comput. **193**, 266 (2007).

[47] Q. Zou and X. He, Phys. Fluids **9**, 1591 (1997).

[48] A. M. Artoli, A. G. Hoekstra and P. M. A. Sloot, Int. J. Mod. Phys. C **13**, 1119 (2002).

[49] A. A. Wheeler, J. Crystal Growth **102**, 691 (1990).

[50] D. Xu, C. Shu, and B. C. Khoo, J. Crystal Growth **173**, 123 (1997).

[51] H. Huang, T. S. Lee, and C. Shu, Int. J. Numer. Methods Fluids **53**, 1707 (2007).

[52] K. Hourigan, L.J.W. Graham, and M.C. Thompson, Phys. Fluid **7,** 3126 (1995).

[53] K. Fujimura, H.S. Koyama, and J.M. Hyun, Trans ASME: J. Fluids Eng. **119**, 450 (1997).

[54] K. Fujimura, H. Yoshizawa, R. Iwatsu, and H.S. Koyama, J. Fluids Eng. **123**, 604 (2001).

[55] A.Y. Gelfgat, J.M. Bar-Yoseph, and A. Solan, J. Fluid Mech. **311**, 1 (1996).

[56] H.M. Blackburn and J.M. Lopez, Phys. Fluids **12**, 2698 (2000).

[57] E. Serre and P. Bontoux, J. Fluid Mech. **459**, 347 (2002).

[58] S. K. Bhaumik and K. N. Lakshmisha, Comput. Fluids **36**, 1163 (2007).

[59] Z. L. Guo, C. Zheng, and B. Shi, Chin. Phys. **11**, 0366 (2002).

[60] K. N. Premnath and S. Banerjee, Phys. Rev. E **80**, 036702 (2009).

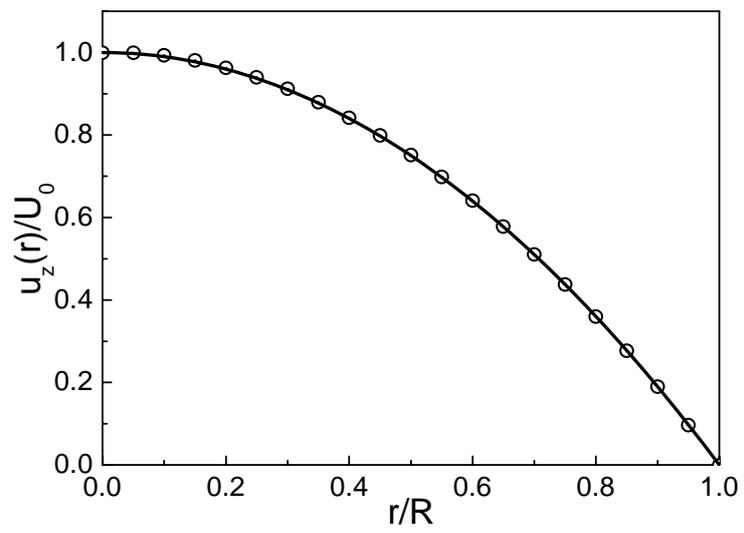

FIG. 1. Analytical (solid line) and numerical (symbol) results of Hagen-Poiseuille flow.

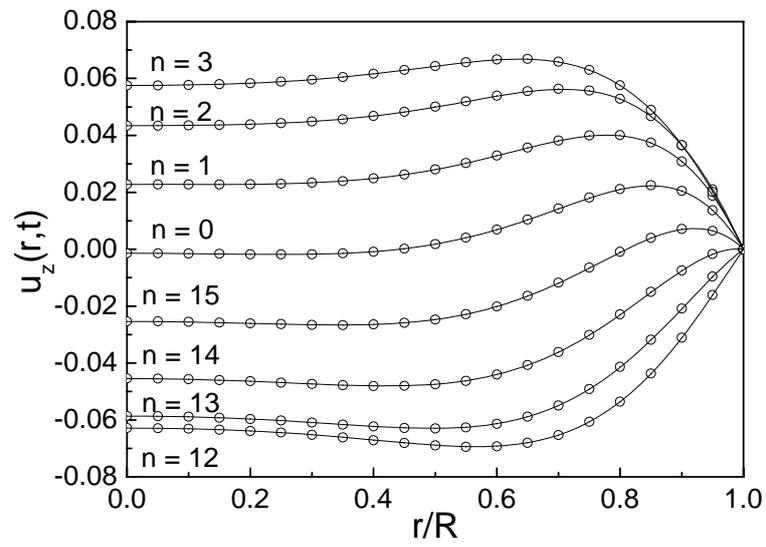

FIG. 2. Analytical (solid line) and numerical (symbol) results of Womersley flow

at different time $t = nT/16$ with $n = 0, 1, 2, 3, 12, 13, 14, 15$.

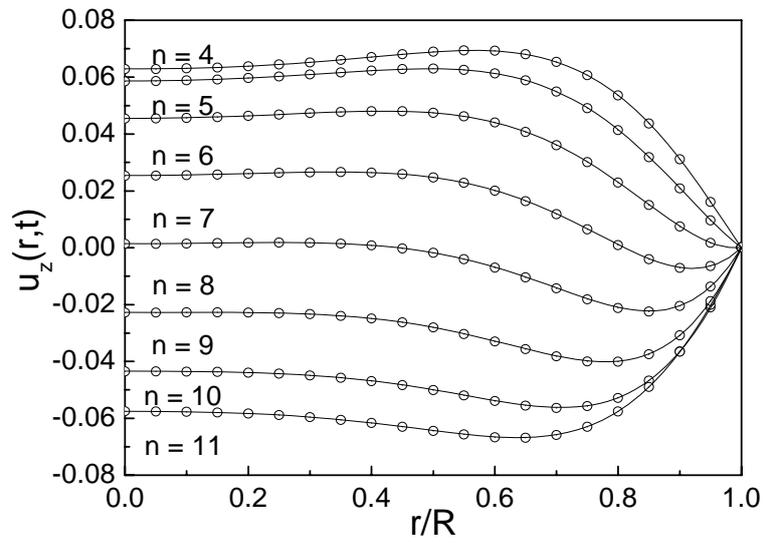

FIG. 3. Analytical (solid line) and numerical (symbol) results of Womersley flow

at different time $t = nT/16$ with $n = 4, 5, 6, 7, 8, 9, 10, 11$.

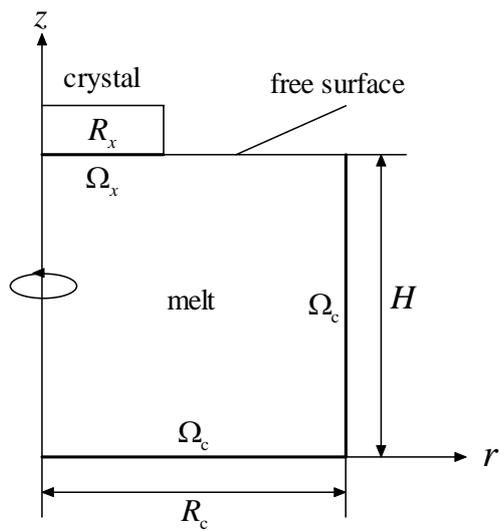

FIG. 4. Configuration of the Wheeler problem.

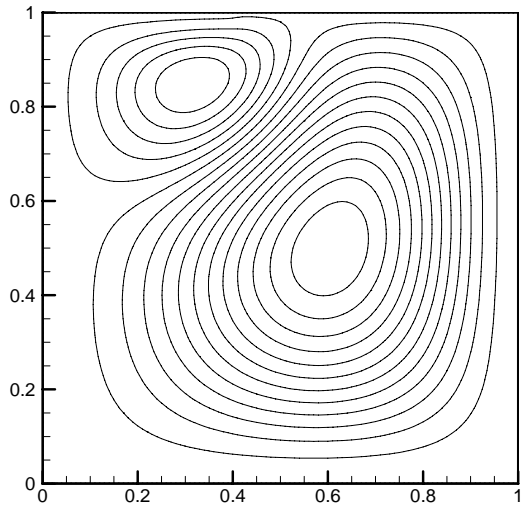

(a)

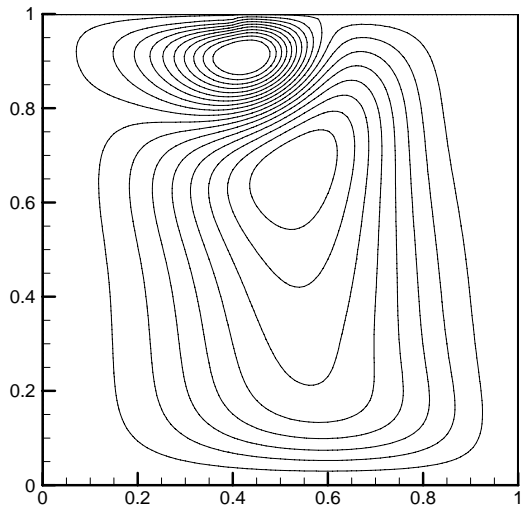

(b)

FIG. 5. Streamlines of the Wheeler problem: (a) $\text{Re}_x = 10^2, \text{Re}_c = -25$; (b) $\text{Re}_x = 10^3, \text{Re}_c = -250$.

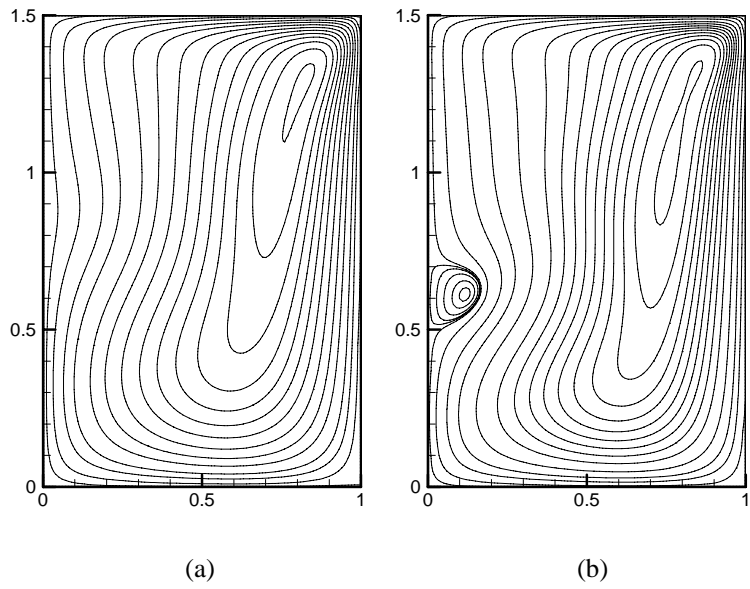

FIG. 6. Streamlines of cylindrical cavity rotational flow: (a) Re = 990 ; (b) Re = 1290 .

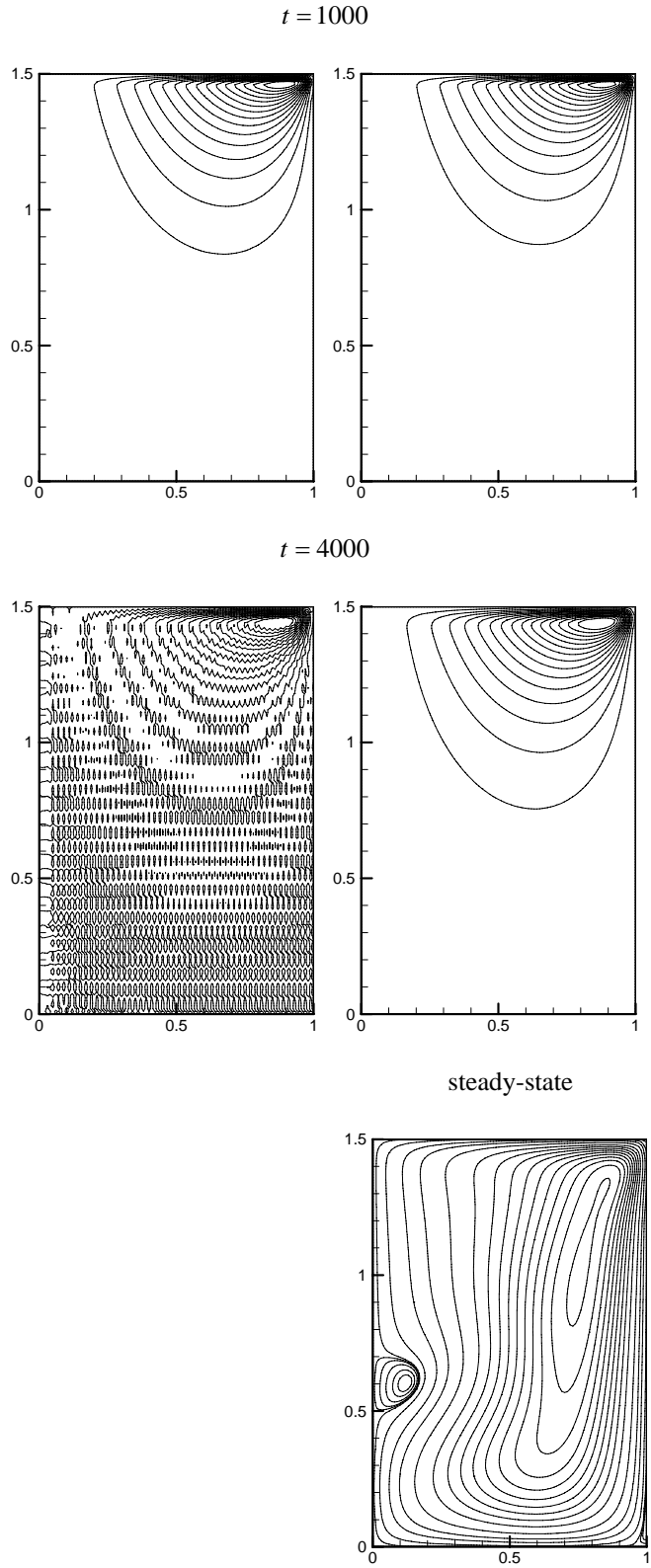

FIG. 7. Simulations of cylindrical cavity rotational flow at $Re = 1290$ with a low viscosity $\nu = 1.67 \times 10^{-3}$ using the BGK (left) and MRT (right) collision models.

Table 1. Comparisons of minimum and maximum stream function for the Wheeler problem.

| Reference | $\text{Re}_x = 10^2, \text{Re}_c = -25$ | | $\text{Re}_x = 10^3, \text{Re}_c = -250$ | |
|---|---|---|---|---|
| | $\psi_{min}$ | $\psi_{max}$ | $\psi_{min}$ | $\psi_{max}$ |
| Present | −0.0494 | 0.1180 | −1.444 | 1.128 |
| Ref. [26] | −0.0514 | 0.1140 | −1.478 | 1.114 |
| Ref. [50] | −0.0443 | 0.1177 | −1.478 | 1.148 |

Table 2. Comparisons of magnitude and location of the maximum axial velocity on the axis for cylindrical cavity rotational flow.

| Reference | Re = 990 | | Re = 1290 | |
|---|---|---|---|---|
| | $u_{z,max}$ | $h_{max}/H$ | $u_{z,max}$ | $h_{max}/H$ |
| Present | 0.0987 | 0.213 | 0.0716 | 0.147 |
| Experimental [54] | 0.097 | 0.21 | 0.068 | 0.14 |
| 3D LB model [58] | 0.093 | 0.22 | 0.072 | 0.16 |